# Size dependent etching of nanodiamond seeds in the early stages of CVD diamond growth

R. Salerno[1,2], B. Pede[1,2], M. Mastellone[2], V. Serpente[2], V. Valentini[2], A. Bellucci[2], D.M. Trucchi[2], F. Domenici[1], M. Tomellini[1*], R. Polini[1,2*]

[1] Dipartimento di Scienze e Tecnologie Chimiche, Università di Roma "Tor Vergata" and Consorzio INSTM RU "Roma Tor Vergata", Via della Ricerca Scientifica 1, Rome, 00133, Italy.

[2] Istituto di Struttura della Materia (ISM), Consiglio Nazionale delle Ricerche (CNR), Sez. Montelibretti, DiaTHEMA Lab, Via Salaria km 29.300, 00015 Monterotondo, Italy.

**Abstract**

We present an experimental study on the etching of detonation nanodiamond (DND) seeds during typical microwave chemical vapor deposition (MWCVD) conditions leading to ultra-thin diamond film formation, which is fundamental for many technological applications. The temporal evolution of the surface density of seeds on Si(100) substrate has been assessed by scanning electron microscopy (SEM). The resulting kinetics have been explained in the framework of a model based on the effect of particle size, according to the Young-Laplace equation, on both chemical potential of carbon atoms in DND and activation energy of reaction. We found that seeds with size smaller than a "critical radius", $r^*$, are etched away while those greater than $r^*$ can grow. Finally, the model allows to estimate the rate coefficients for growth and etching from the experimental kinetics.

* Corresponding authors.

*E-mail addresses*: tomellini@uniroma2.it (M. Tomellini); polini@uniroma2.it (R. Polini)

# 1  Introduction

Diamond is an intriguing material for applications in materials science, engineering, chemistry, and bioscience because of the unique combination of outstanding and tunable properties [1,2]. In particular, nanocrystalline diamond (NCD) is used as a functional coating in many applications [2-6] and is an emerging material for quantum information technologies [7] and advanced biolabeling [8].

It is a well-established fact that diamond films can be grown by chemical vapor deposition (CVD) using carbon-containing gas species such as methane, aliphatic and aromatic hydrocarbons, alcohols, ketones, amines, ethers and carbon monoxide. Methane diluted in hydrogen is the most common feed gas employed at both lab and industrial scale.

Due to its large surface energy, diamond grows on heterosubstrates by the Volmer-Weber mechanism, i.e., heterogeneous nucleation and island growth at the substrate surface. However, in most CVD methods the nucleation density of diamond islands on non-diamond surfaces is usually very low, typically in the range $10^4$–$10^6$ nuclei/cm² [9,10]. This implies that only rather thick films ($> 10^1$–$10^2$ μm) can be formed by coalescence of the growing nuclei/islands. To get ultrathin ($\leq$ 100 nm) NCD layers, nucleation densities exceeding $10^{11}$ cm$^{-2}$ ($10^3$ μm$^{-2}$) are mandatory [11-13]. Therefore, several nucleation enhancement methods, including abrading with diamond grit and negative biasing of the substrate surface, have been developed to increase the nucleation densities [14,15]. Another way to get ultrathin films is the seeding technique, i.e., the application of tiny diamond particles at the substrate surface, which act as growing centers in the subsequent CVD process [11]. The market availability of detonation nanodiamond (DND) suspensions enables the application of a large number of diamond nanoparticles per unit area, with good spatial homogeneity and size distribution. When the substrate is sonicated in a suspension of deagglomerated DNDs with average diameter around 4-5 nm, the interaction between DNDs and the substrate surface leaves seed densities as large as $10^{12}$ cm$^{-2}$ [16]. The further reduction of DNDs' average size allows to attain even larger seeds' densities ($10^{13}$ cm$^{-2}$) and the growth of diamond films as thin as 5.5 nm [17,18] if proper CVD conditions are employed. S. Stehlik and coworkers wrote in [18]: "A high concentration of atomic hydrogen might also lead to a certain reduction of the nucleation[1] density due to the possible etching of nanodiamond seeds. To suppress these effects, we performed the NCD growth at lower pressure and lower temperature". The

---

[1] Several Authors in the relevant literature refer to seeds as nuclei; consequently, the terms "seeding density" and "nucleation density" are used as synonyms. We prefer to call "nuclei" those crystals that form because of a nucleation process. Diamond seeds deposited at the substrate surface after a sonication process, as well as diamond fragments that scratching pretreatments could implant in the substrate surface, should be considered as "growth centers", not as "nuclei", in that they do not originate at the solid-gas interface via a heterogeneous nucleation process during CVD.

Authors had to use a Linear Antenna Microwave (MW) plasma system ensuring both lower substrate temperature (460 °C) and lower pressure (0.11 Torr) for the growth of the extremely small (~2 nm) seeds they were able to prepare and disperse at the substrate surface. Under those specific CVD conditions, the atomic hydrogen concentration in the activated gas mixture is expected to be much lower than that typically achieved in conventional Microwave (MW) CVD reactors; the deposition rate was therefore also significantly lower ($\approx$ 1 nm/h). A lower deposition temperature, in combination with a low monohydrogen density in the plasma, was necessary for the survival of the 2 nm nanodiamonds during the diamond growth process. In fact, the etching of diamond in hydrogen microwave plasma is a thermally activated process, with activation energies in the interval 32-45 kcal/mol [19,20].

Evan L.H. Thomas et al. [21] mentioned that "After 4 min of growth … reduction in the density of crystallites is observed". Practically, the Authors noted a reduction of the initial surface density of ND seeds even if they tried to avoid the phenomenon, as they wrote in the Experimental Methods section: "During CVD, a methane fraction of 3.86% diluted in hydrogen was used to stabilize and prevent etching of the ~5 nm seeds for the first 3 min, before being reduced to 0.6% for the remainder of growth".

It is evident that DND seeds may not be stable under diamond MW CVD conditions.

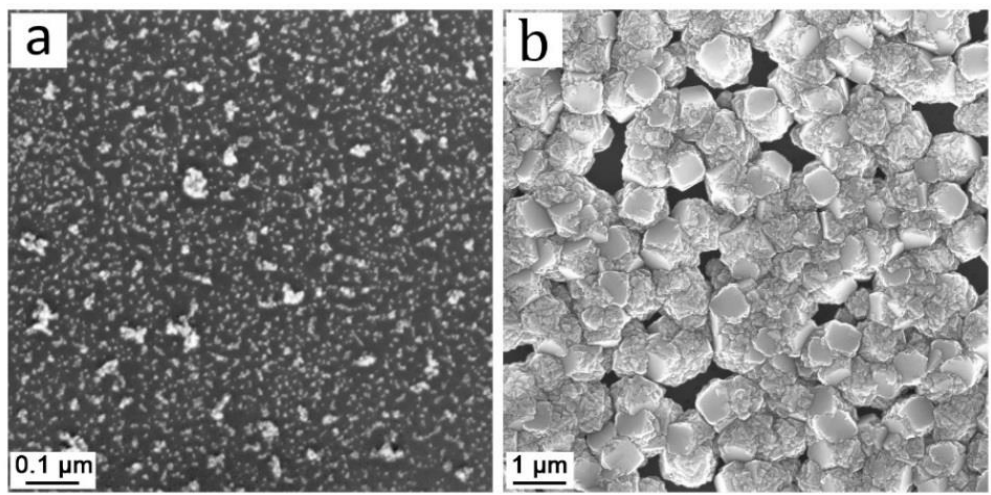

**Fig. 1** Si(100) substrate seeded using DND *(a)*, 0.4 µm diamond coating deposited by MWCVD on the same seeded substrate *(b)*. If all seeds were grown, the deposit should have been continuous.

In our lab experience, we have also noted that a fraction of DND seeds does disappear under MWCVD conditions suitable for diamond film growth. Figure 1 shows an as-seeded Si(100) substrate (panel *a*, seed density > $10^{11}$ cm$^{-2}$) and the same substrate after 50 min MWCVD (840 °C, MW power = 1.25 kW, *p* = 40 Torr). The deposit of Fig. 1*b* is not continuous, and its average thickness was ~0.4 µm. Considering that a seeding density exceeding $10^{11}$ cm$^{-2}$ should ensure the formation of a continuous film thinner than 100 nm [13], it follows that a fraction of the initial population of seeds did not grow.

The aim of this paper is to study the kinetics of seed disappearance and to provide an explanation of this phenomenon, which – in our knowledge – has not been thoroughly investigated so far. We propose a phenomenological model for describing the behavior of the experimental kinetics of seed disappearance from the surface. The model predicts that ND etching is a particle-size dependent process, which is more effective the lower the size of the seed.

The article is divided according to the following. In section 2, we describe the experiments of diamond growth by seeding, the characterization of the film and the procedure employed to obtain kinetic data on the etching process. In section 3, the kinetic data are described in the framework of a kinetic model for size-dependent stability of ND. The modeling allows us to gain information on the rate coefficients for both growth and etching process.

## 2  Experimental

### 2.1  Seeding and deposition

The samples used in this work were obtained from 10 cm diameter, double polished silicon wafers (270 ± 25 µm thickness, (100) orientation, p-type doped with boron, resistivity of 1–5 mΩ·cm, surface roughness lower than 1 nm) supplied by University Wafer Inc. The wafers were cut down to 1.5cm × 1cm samples that underwent a cleaning and seeding process. The cleaning comprised 10 min ultrasonic bath in acetone, 10 min ultrasonic bath in 2-propanol and 3 min ultrasonic bath in a HF:H$_2$O = 1:9 solution followed by an abundant rinse with deionised water. Each step was followed by drying with compressed air flow. After the cleaning steps, the samples were seeded with a detonation nanodiamond (DND) suspension obtained by mixing commercially available DND suspension in Dimethyl sulfoxide (*Blueseeds*, by Adámas Nanotechnology, 0.5% w/v ND in DMSO) and methanol in 1:3 ratio [22]. The substrates were immersed in the DND suspension and sonicated for 15 minutes in ultrasonic bath. Finally, the substrates were rinsed with methanol and dried under compressed air flow. An average seeding density of (7 ± 2) × $10^{11}$ cm$^{-2}$ was obtained after the process (Fig. 2, top panel). In order to determine the DND particle density before and after each CVD process, the seeded substrates were delicately cleaved: one third of the substrate was set aside in order to measure the surface density of seeds by SEM; the remaining 1cm × 1cm sample was loaded into the CVD chamber.

Diamond growth was accomplished via a customized microwave-plasma assisted CVD ASTeX reactor equipped with a MKS Instruments TM025 51 microwave generator operating at 2.45 GHz frequency. After placing the substrate on top of a molybdenum disc positioned on the graphite stage in the reaction chamber, the pressure was lowered to $2\times10^{-6}$ Torr at which point the temperature was gradually raised to 700 °C while keeping the pressure below $9\times10^{-6}$ Torr. Once the base pressure returned to $2\times10^{-6}$ Torr, hydrogen and methane were introduced in the reaction chamber. The methane concentration was either 0.5 vol.% or 1 vol.%. The total gas flow was 200 standard cubic centimetre per minute (sccm). Microwave power used in this work was 300 W, 600 W or 800 W; to guarantee a stable plasma discharge in the MW CVD reactor, we used 8 Torr, 24.5 Torr and 27 Torr deposition pressure, respectively. The deposition time was counted starting when the plasma was ignited and ended when the plasma was shut down. After shutting down the plasma, the methane flow was cut off and the temperature was gradually lowered back to room temperature.

*2.2 Characterization*

The DND particle size distribution in the suspension employed for substrate seeding was assessed by photon correlation spectroscopy (PCS) using a Malvern Instrument Zetasizer Nano ZS equipped with a 4 mW helium-neon laser (632.8 nm); the scattering angle was 173°. A refractive index of 1.352 and a viscosity equal to 0.722 mPa·s at 20 °C of the DND dispersion medium (1 part DMSO and 3 parts methanol) were used to convert the measured intensity/size distribution to volume/size distribution ($f_v(r)$; see Fig. 2, bottom panel).

Field emission gun scanning electron microscopy (FEG SEM, Zeiss LEO Supra 35) was used to determine seeding density, as well as surface morphology and thickness of the deposited NCD films. All SEM images were acquired at 10 kV in the regime of secondary electrons (SEs), using the Inlens SE detector of the microscope. SEM micrographs were analyzed by ImageJ software to estimate the surface density of diamond seeds/crystallites.

Raman spectroscopy measurements were carried out using a Horiba Scientific LabRam HR Evolution confocal spectrometer equipped with a 100 mW laser source (wavelength of 532 nm, by Oxxius SA, France); a computerized XY-table, an electron-multiplier CCD detector, an Olympus U5RE2 microscope with 100× objective (laser spot on the sample surface 0.7 μm) with a numerical aperture (NA) of 0.9, and a grating with 600 grooves/mm were used. All Raman spectra were recorded in backscattering geometry focalizing 100 mW at the sample surface; twenty spectra with an accumulation time of 10 s were averaged.

Numerical computations have been performed using the Wolfram Mathematica package.

## 3 Results and discussion

### 3.1 Size distribution of the DND suspension

Fig. 2 illustrates the results of DND seeding with small average aggregate size. PCS results were in good agreement with FEG SEM characterization. The $f_v(r)$ distribution function (red dotted line) was bimodal, with two modes at 2.8 nm and 9 nm of radius. The second peak in the probability density function accounts for 34% of total volume. These findings are in excellent agreement with previous results published by O. Shenderova et al., who used the same suspension (Fig. 3 of ref. [22]). The $f(r)$ distribution function (black line in Fig. 2), i.e., the fraction of particles ($f(r)dr$) with radius in the range $(r, r + dr)$ at $r$, was derived from the $f_v(r)$, as detailed in the Supplementary Material. The second mode of the $f_v(r)$ is not clearly visible anymore in the size distribution function, $f(r)$.

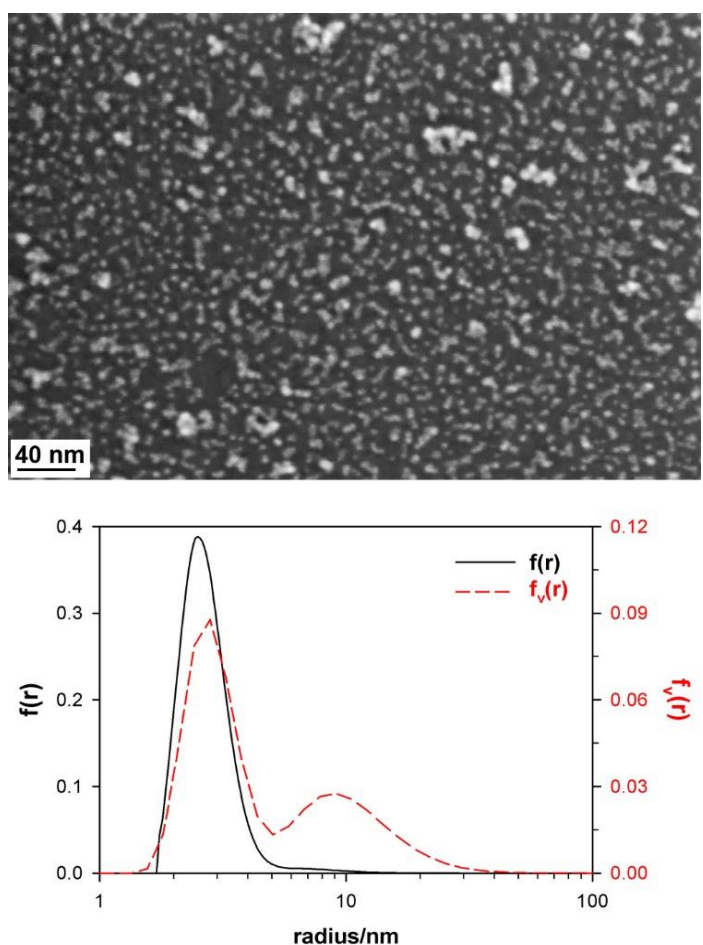

**Fig. 2** SEM micrograph of DNDs seeded on Si(100) substrate by treating it in an ultrasonic bath with 1 part of 0.5% w/v DND in DMSO and 3 parts of $CH_3OH$. The plot shows the size distribution by volume,

$f_v(r)$, (red dotted line), and the size distribution by number, $f(r)$ (black line), of the suspension used for the seeding. The $f_v(r)$ was obtained by PCS; the $f(r)$ was calculated from the $f_v(r)$.

*3.2 Stability of DND seeds under high vacuum annealing and $H_2$ plasma exposure*

As pointed out in the Introduction, preliminary results had provided indirect evidence that a fraction of initial seeds was missing after CVD (see Fig. 1). The seeds could have disappeared during diamond CVD for two reasons: *a)* etching/gasification performed by the activated gas phase, *b)* high temperature solid state reaction between sp$^3$ carbon and silicon substrate. To check if ND seeds were thermally stable at the deposition temperature, we subjected as-seeded Si(100) to a prolonged annealing (48 h at 700 °C) in the MW CVD reactor under high vacuum ($p = 2 \times 10^{-7}$ Torr). Fig. 3 displays the SEM micrographs of the as-seeded substrate (panel *a*) and of the same sample after annealing (panel *b*). The seed density did not change, thus indicating that DND particles deposited at the Si(100) surface are stable at 700 °C. These findings confirm previously published results [23].

We also checked the stability of DND seeds under $H_2$ plasma exposure in our MWCVD reactor, without methane in the feed gas. Panel *c* of Fig. 3 displays the Si(100) substrate surface after 10 min of $H_2$ exposure at 700 °C and 600 W microwave power ($p = 24$ Torr).

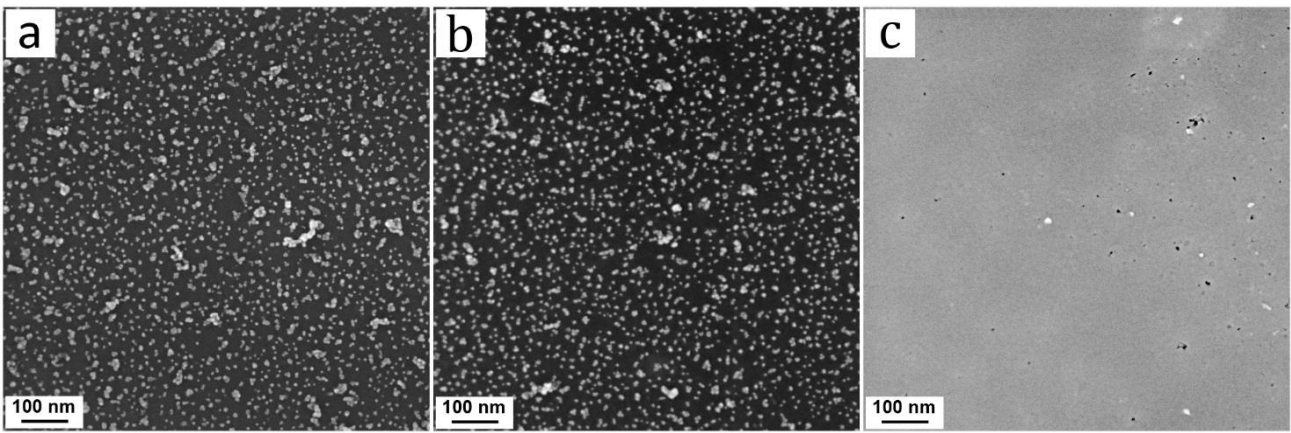

**Fig. 3** SEM micrographs of DNDs seeded on Si(100) substrate *(a)*, after 48 h annealing at 700 °C in high vacuum *(b)*, and after 10 min $H_2$ plasma exposure at 700 °C, 600 W, 24 Torr *(c)*.

It is evident that almost all of the seeds have disappeared, with just few particles left at the Si(100) surface. The bright particles visible in Fig. 3, panel *c*, represent what remained of the largest agglomerates initially present. If we consider that most of the seeds (≥ 99.9%) were etched away, we have calculated, as discussed in § 3.4, that the rate of etching under pure $H_2$ plasma exposure was at least 0.4 nm/min (≥ 24 nm/h). J.C. Arnault and coworkers [24] estimated a nanodiamond etching rate lower than 4.4 nm/h under $H_2$ plasma at even higher temperature ($T = 940$ °C). The nearly one order of magnitude larger etching rate we have observed at 700 °C might be ascribed to a larger plasma density and/or to a higher atomic hydrogen concentration at the substrate surface in our reactor.

Atomic hydrogen not only eliminated the DND particles, but caused the etching of Si(100) resulting in the formation of the dark etch pits detected by SEM (Fig. 3 *c*) [25].

Therefore, our preliminary seed stability scouting investigation demonstrated that *i)* DND seeds on Si(100) are thermally stable at the CVD temperature we employed for studying seeds' density evolution under diamond CVD conditions, *ii)* DND seeds are quickly gasified when exposed to pure $H_2$ plasma in our MWCVD reactor.

### *3.3 Stability of DND seeds under diamond MWCVD*

The aim of this work was to study the (in)stability of DND particles acting as growth centers in the deposition of NCD films. We performed short deposition runs to study the time evolution of DND seeds at the substrate surface, by selecting different deposition conditions that allow the formation of diamond films.

| MW plasma power (W) | $CH_4$ % | $p$ (Torr) | CVD duration ($t$, min) | $N(0)/cm^{-2}$ | $N(t)/cm^{-2}$ | $N\%$ |
|---|---|---|---|---|---|---|
| 300 | 0.5 | 8 | 3 | $(7.9 \pm 0.4) \times 10^{11}$ | $(5.3 \pm 0.4) \times 10^{11}$ | $67 \pm 6$ |
| | | | 5 | $(8.0 \pm 0.4) \times 10^{11}$ | $(4.8 \pm 0.3) \times 10^{11}$ | $60 \pm 5$ |
| | | | 10 | $(7.2 \pm 0.5) \times 10^{11}$ | $(3.7 \pm 0.3) \times 10^{11}$ | $51 \pm 5$ |
| | | | 40 | $(10.6 \pm 0.2) \times 10^{11}$ | $(1.6 \pm 0.2) \times 10^{11}$ | $15 \pm 2$ |
| 600 | 0.5 | 24.5 | 3 | $(8.2 \pm 0.1) \times 10^{11}$ | $(5.3 \pm 0.2) \times 10^{11}$ | $65 \pm 2$ |
| | | | 5 | $(6.3 \pm 0.2) \times 10^{11}$ | $(3.5 \pm 0.1) \times 10^{11}$ | $56 \pm 2$ |
| | | | 10 | $(7.7 \pm 0.6) \times 10^{11}$ | $(3.1 \pm 0.2) \times 10^{11}$ | $40 \pm 4$ |
| 600 | 1 | 24.5 | 2 | $(8.0 \pm 0.3) \times 10^{11}$ | $(5.5 \pm 0.2) \times 10^{11}$ | $69 \pm 4$ |
| | | | 3 | $(5.5 \pm 0.1) \times 10^{11}$ | $(3.6 \pm 0.1) \times 10^{11}$ | $65 \pm 2$ |
| | | | 5 | $(6.3 \pm 0.2) \times 10^{11}$ | $(3.5 \pm 0.1) \times 10^{11}$ | $56 \pm 2$ |
| | | | 7 | $(7.8 \pm 0.1) \times 10^{11}$ | $(2.5 \pm 0.2) \times 10^{11}$ | $32 \pm 3$ |
| 800 | 0.5 | 27 | 2 | $(9.0 \pm 0.3) \times 10^{11}$ | $(6.3 \pm 0.1) \times 10^{11}$ | $70 \pm 3$ |
| | | | 3 | $(6.9 \pm 0.3) \times 10^{11}$ | $(4.2 \pm 0.1) \times 10^{11}$ | $61 \pm 3$ |
| | | | 5 | $(8.9 \pm 0.2) \times 10^{11}$ | $(4.8 \pm 0.3) \times 10^{11}$ | $54 \pm 4$ |

**Table 1** DND seeds' evolution after short MWCVD deposition runs; substrate temperature was 700 °C.

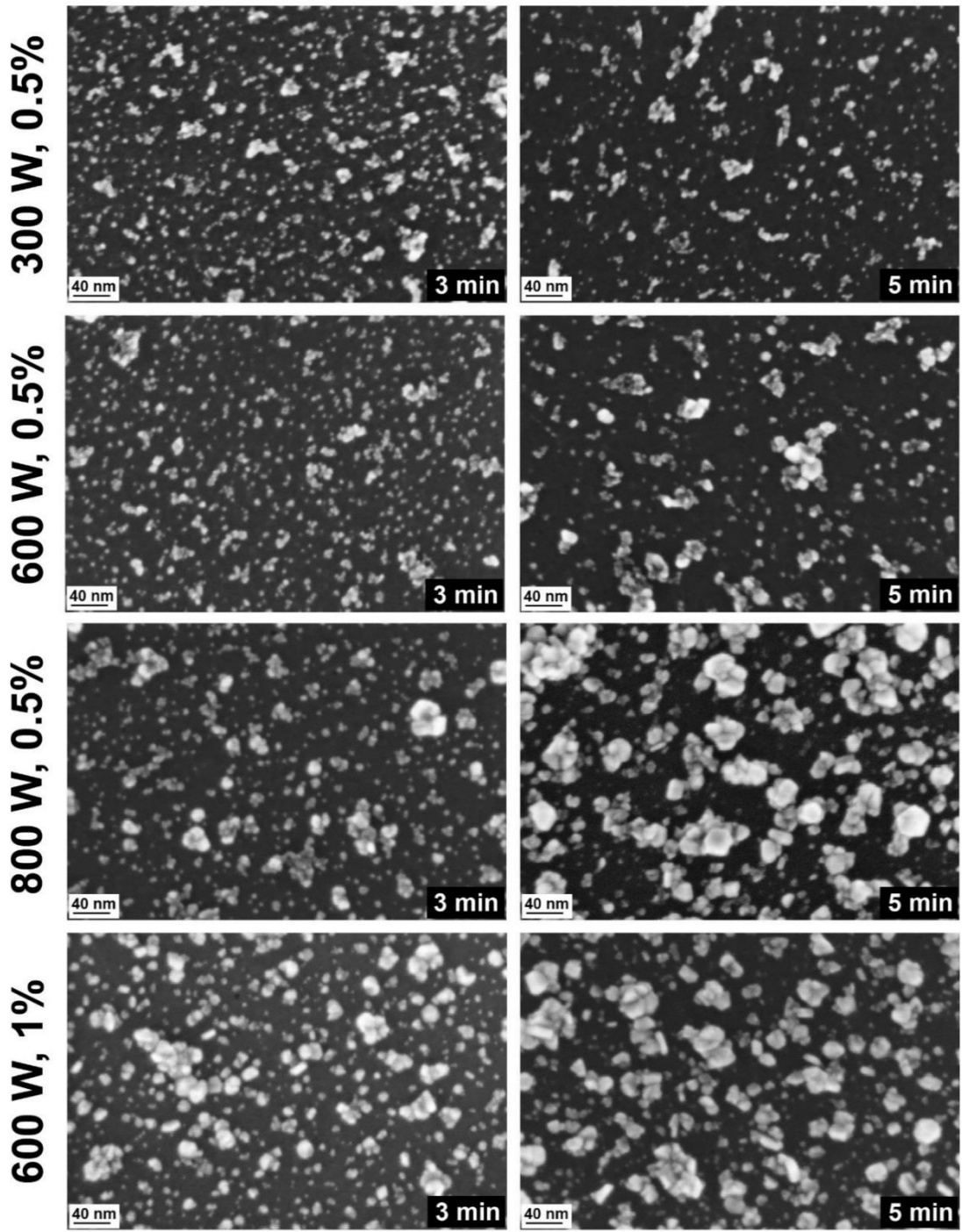

**Fig. 4** SEM micrographs of nanodiamond crystallites after 3- and 5-min deposition runs (see Table 1 for surface densities).

Table 1 summarizes the depositions we performed leading to non-continuous deposits. $N(0)$ represents the initial seed density for a given sample, $N(t)$ is the density of crystallites after $t$ min CVD, $N\% = 100 \times N(t)/N(0)$ is the percentage of crystallites remaining on the substrate surface after $t$ min CVD. For each sample we used five SEM micrographs to determine $N(0)$ and $N(t)$ by image analysis.

With the only exception of data at 300 W (0.5% CH$_4$), it was not possible to get meaningful $N(t)$ values for CVD times longer than 10 min, due to the quick coverage of the substrate surface caused by growing seeds.

Fig. 4 shows the SEM micrographs of samples subjected to 3- and 5-min CVD. It is worth comparing those micrographs to the SEM image of Fig. 2, showing the larger DND surface density ($N\% = 100$) of a typical as-seeded substrate.

In Table 2 the CVD durations at film coalescence are reported, as well as thickness and deposition rate of the continuous films. The morphology and Raman spectra of these films are displayed in Figs. S1 and S2 of Supplementary Material, respectively.

| MW plasma power (W) | CH$_4$ % | $p$ (Torr) | time at coalescence ($t$, min) | film thickness (nm) | deposition rate (nm/min) |
|---|---|---|---|---|---|
| 300 | 0.5 | 8 | - | - | - |
| 600 | 0.5 | 24.5 | 20 | 39 ± 4 | 1.95 ± 0.2 |
| 600 | 1 | 24.5 | 12 | 45 ± 11 | 3.75 ± 0.9 |
| 800 | 0.5 | 27 | 15 | 52 ± 10 | 3.5 ± 0.7 |

**Table 2** Thickness of the continuous NCD films at coalescence and corresponding deposition conditions and growth rates.

The data show that, by using 600 W and 800 W microwave (MW) power, the surface density of initial seeds decreased (see Table 1) while some seeds did actually grow (Fig. 4). A different scenario occurred at 300 W, where a significant crystallite growth did not occur even after 40 min, although the surface density of seeds continued to decrease with CVD time (see Fig. S3).

To check if diamond could grow by using such a low plasma power, we put in the CVD chamber the sample previously coated with 52 nm continuous film deposited for 15 min at 800 W, 0.5% CH$_4$ (Table 2). This sample was subjected to a second CVD run for 300 min CVD, at 300 W, 0.5% CH$_4$. The thickness of the previously prepared NCD film augmented from 52 nm to 180 nm. This experimental result indicates that when a continuous NCD film was exposed to the same MWCVD conditions that caused seeds disappearance, its thickness did increase, with a rate – in this specific case – of 0.43 nm/min: the growth rate of "flat" diamond is larger than that of the seeds. Similar results were reported by A. Giussani and coworkers [26]; these Authors used a O$_2$ containing feed gas in the MWCVD reactor (1% CH$_4$, 1% O$_2$, 98% H$_2$, Si(100) substrate, T = 798 ± 7 °C): by employing the same MWCVD conditions which caused all seeds to disappear on Si(100), they could increase the thickness of a diamond film previously prepared.

Consequently, considering that DND seeds on Si(100) are thermally stable at 700 °C, the decrease of the surface density of nanodiamonds must be ascribed to the etching of some seeds, while some other seeds can survive and/or act as growth centers till the coalescence of the film. A spontaneous question arises at this point: which seeds are disappearing, and why?

*3.4    Kinetic approach for the size-dependent stability of DND*

In this section we present a kinetic model for the etching of ND particles, by taking into account the effect of seed size on kinetics. The etching mechanism of diamond includes numerous surface chemical reactions, but – as was shown in [27] – only few of them play a principal role in the process. Therefore, we consider here the following schematic reaction for the etching of nanodiamond:

$$C_{(s)} + x\, H_{(g)} \rightarrow CH_{x\,(g)}, \qquad (1)$$

where $C_{(s)}$ stands for sp³ carbon atoms in the ND, eventually bonded to adsorbed hydrogen atoms[2], and $H_{(g)}$ is atomic hydrogen in the plasma. Seeds can be grouped in dependence of their volume, say $v_0$, from which an equivalent radius, $r_0$, is defined: $r_0 = \left(\frac{3}{4\pi} v_0\right)^{1/3}$. We stress that, owing to the detonation production process of DND, seeds with the same volume (or equivalent radius) could have, in general, different shapes, or crystal habits [28]. Since the etching rate is expected to depend upon the habit, in the following the etching rate of the set of seeds with size $r_0$ is to be intended in the statistical sense, i.e., as an average quantity.

The rate of the etching reaction (moles/s) is a function of temperature and gas phase composition (that is considered to be time independent) according to $\left(\frac{dn}{dt}\right)_e = K'_e(T,r) \cdot s$ where $s = 4\pi r^2$ is the surface of the seed particle, assumed spherical, $r$ its radius, and $K'_e$ the rate coefficient of the forward reaction (1). In particular, $K'_e(T,r) = v e^{-\Delta G^{\#}_e(r,T)/RT}$ with $v$ the pre-exponential factor and $\Delta G^{\#}_e(r,T)$ the activation free energy depending on gas phase composition and seed size. This last dependence is ascribed to the size dependence of the chemical potential of the solid reactant, namely the sp³ C atoms, that is related to the surface tension of the crystal. The increment of the chemical potential of C in the ND particles is due to the increase of stress (pressure) as given by the Young-Laplace equation. This contribution does reduce the energy barrier according to $\Delta G_e^\#(r,T) = \Delta g^\#(T) - \frac{2\sigma v_m}{r}$, where $\sigma$ is the surface tension of diamond, $v_m$ the molar volume and $\Delta g^\#(T)$ the activation energy in the absence of the Young-Laplace (YL) term. It is worth noting that the YL equation plays a key role in describing the thermodynamics of systems at the nanoscale [29]. In addition, the YL equation has been used to extrapolate the diamond-graphite P-T equilibrium phase diagram to the nanometer-scale region [30].

---

[2] We do not consider intentionally the sp² shell that surrounds DND particles in that it is supposed to be quickly etched away in the presence of atomic hydrogen in the CVD reactor.

Alternatively, we can attain a similar result by exploiting the detailed balancing for the reaction above, according to: $\frac{K_e^{(f)}}{K_e^{(b)}} = e^{-\Delta G^0/RT}$ with $K_e^{(b)}$ and $K_e^{(f)}$ rate constants of the backward and forward reactions, respectively, and $\Delta G^0$ the standard free energy change of reaction (1). It stems that $(\Delta G^0(r) - \Delta G^0(\infty)) = -\frac{2\sigma v_m}{r}$, that leads to $K_e^{(f)}(r) = K_e^{(f)}(\infty)e^{\frac{2\sigma v_m}{rRT}} = v e^{\frac{\Delta g^{\#(0)}(T)}{RT}} e^{\frac{2\sigma v_m}{rRT}}$, where the rate constant of the reverse reaction has been considered independent of $r$.

On the other hand, the growth reaction is also proportional to the particle surface area (i.e., to the number of adsorption site [31]) as $\left(\frac{dn}{dt}\right)_g = K_g(T) \cdot s$ where $K_g(T) = v e^{-\Delta G^\#(T)/RT}$, with size independent activation energy. A pictorial view of the effect of particle size on energy barrier is displayed in Fig. 5. In the figure we also indicate the reaction affinity, i.e., $A = \sum_i v_i \mu_i$, with $v_i$ stoichiometric coefficients (positive and negative for products and reactants, respectively) and $\mu_i$ chemical potentials. Specifically, the chemical potentials are $\mu_C(r) = \mu_C(\infty) + \frac{2\sigma v_m}{r}$, $\mu_H = \mu_H^0 + RT \ln P_H/P^0$ and $\mu_{CH_x} = \mu_{CH_x}^0 + RT \ln P_{CH_x}/P^0$.

The net rate of growth is therefore given by $\left(\frac{dn}{dt}\right) = -K'_e(T,r) \cdot s + K_g(T) \cdot s$, that is ($n = \frac{4\pi r^3}{3v_m}$, $s = 4\pi r^2$):

$$\frac{dr}{dt} = -K_e e^{c/r} + K_g \tag{2}$$

where $K_e = v e^{-\Delta g^\#(T)/RT}$ and $c = 2\sigma v_m/RT$. Integration of eqn. 2 provides the $r(r_0, t)$ function, that is here intended as the evolution of the average radius of the $r_0$-population of seeds. In other words, as anticipated above, owing to the polydispersivity of diamond nanoparticles and the dependence of the rate of etching on crystal face $r$ and $K_e$ are considered average quantities.

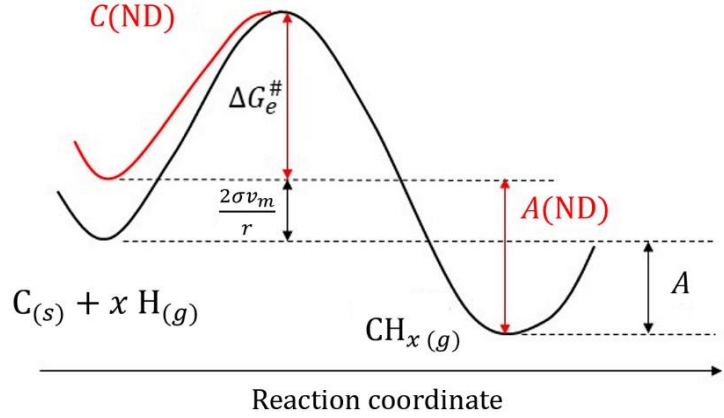

**Fig. 5** Pictorial view of the free energy profile for nanodiamond particles etching reaction. The free energy curves, with and without the inclusion of the curvature effect, are displayed as red and black lines, respectively. In the figure, $A$ is the affinity of the etching reaction, that is a decreasing function of the radius of the ND particle.

By denoting with $k_g = K_g - K_e$ the net rate of ND growth in the absence of the YL term, the rate equation eqn. 2 can be rewritten in the form:

$$\frac{dr}{d\tau} = \frac{k_g}{|k_g|} \frac{\exp(c/r^*) - \exp(c/r)}{\exp(c/r^*) - 1}, \qquad (3)$$

where $\tau = t|k_g|$ is the reduced time (dimension of a length) and the parameter $r^* = \frac{c}{\ln(K_g/K_e)} = \frac{c}{\ln(1 + k_g/K_e)}$ has been defined. In eqn. 3, $r^*$ is positive-definite for $k_g > 0$, namely $K_g > K_e$. On the other hand, for $k_g < 0$, i.e., $K_g < K_e$, $r^*$ is negative-definite and $\frac{dr}{d\tau}$ is lower than zero for any value of the seed size; in this latter case all ND particles undergo an etching process and are removed from the substrate.

Eqn. 3 shows that, for $r_0 < r^*$, ND particles shrink and are then etched away completely from the substrate. The opposite occurs for larger seeds ($r_0 > r^*$) where the growth prevails. A set of typical solutions of eqn. 3 for initial sizes of seeds larger and shorter than $r^*$ are reported in Fig. 6. According to eqn. 3, the seeds with initial radius close to $r^*$ will grow at a very low rate.

It is worth noting that a slightly different approach, based on the *kinetic law of mass action* [32, 33], actually leads to an equation similar to eqn. 2.

In the following, we briefly summarize the method employed for modeling the time dependence of the number density of seeds, $N(t)$, during the CVD treatment.

For given $r^*$ and $c$, the numerical integration of eqn. 3 allows one to compute the time dependence of the mean size of the seed population with initial size $r_0$: $r = r(r_0, \tau)$. For a positive value of $r^*$ and $r_0 < r^*$, the mean radius vanishes at a certain time, say $\tau = \tau_s$, that implies $r(r_0, \tau_s) = 0$. In this last equation $\tau_s$ has the meaning of mean-lifetime of the seeds with initial size $r_0$. Explicit form of this relation in the end provides the function $\tau_s = \tau_s(r_0)$.

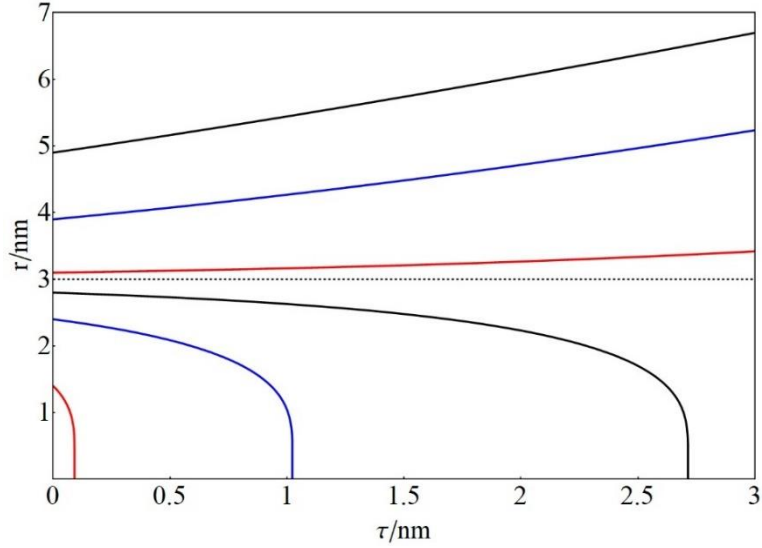

**Fig. 6** Numerical solution of the kinetics of growth and shrinkage for ND seeds having $r_0 > r^*$ and $r_0 < r^*$, respectively. Computations have been done for $r^* = 3$ nm and $c = 3.06$ nm. $\tau$ is the reduced time (see text).

Based on the statistical significance of $\tau_s$ it is possible to include statistical fluctuations of crystal habit into the kinetic model. For a given set of seeds with mean-lifetime $\tau_s(r_0)$, Poisson statistic provides $\frac{dn}{d\tau} = -\frac{n}{\tau_s(r_0)}$, where $n$ is the surface density of the considered set of the $r_0$-ensemble of seeds. Integration of the equation yields

$$n(r_0, \tau) = n(r_0, 0) e^{-\frac{\tau}{\tau_s(r_0)}}, \tag{4}$$

where $n(r_0, 0)$ is the initial density of seeds ($\tau = 0$).

To determine the kinetics of the number density of seeds we resort to the probability density function (PDF) of the deposited ND, $f(r_0)$. As anticipated, the experimental PDF is well described by a log-normal distribution. Specifically, the experimental PDF, as a function of particle radius, is given by the log-normal with mean and variance equal to $m = 0.996$ and $v = 0.22$, respectively, as obtained by fitting procedure. At reduced time $\tau$ the number of seeds belonging to the $r_0$-population, still present on the surface, are given by eqn. 4 in terms of $f(r_0)$ according to

$$dn(r_0, \tau) = N(0) e^{-\frac{\tau}{\tau_s(r_0)}} f(r_0) dr_0, \tag{5}$$

where $N(0)$ is the initial surface density of seeds. The kinetics of the number of etched seeds is given by integrating eqn. 5: $N_e(\tau) = \int_0^{r^*} dn(r_0, \tau)$ while the number of growing seeds, with $r > r^*$, equals $N_g(\tau) = N(0) \int_{r^*}^{\infty} f(x)dx$. The total number of seeds as a function of reduced time is therefore given by $N(\tau) = N_g(\tau) + N_e(\tau)$, namely

$$\frac{N(\tau)}{N(0)} = 1 - \int_0^{r^*} f(x)dx + \int_0^{r^*} e^{-\frac{\tau}{\tau_s(x)}} f(x)dx = 1 - \int_0^{r^*} f(x)\left(1 - e^{-\frac{\tau}{\tau_s(x)}}\right)dx. \quad (6)$$

The value of the cut off radius, $r^*$, can be estimated from the experimental $\frac{N(t)}{N(0)}$ curve at longer time, say $\tau^*$, when $N(t)$ reaches a stationary value. This figure is representative of the total number of seeds surviving on the surface, which contribute to the film growth. By denoting with $\theta = \frac{N(\tau^*)}{N(0)}$ the steady state value of seeds, eqn. 6 provides $\theta = \int_{r^*}^{\infty} f(x)dx$ from which $r^*$ can be estimated. The knowledge of $r^*$ allows us to integrate eqn. 3 and to determine the $\tau_s(r_0)$ function which enters eqn. 6.

In Fig. 7 the results of the modeling are compared to the experimental data-points (Table 1) for several deposition conditions.

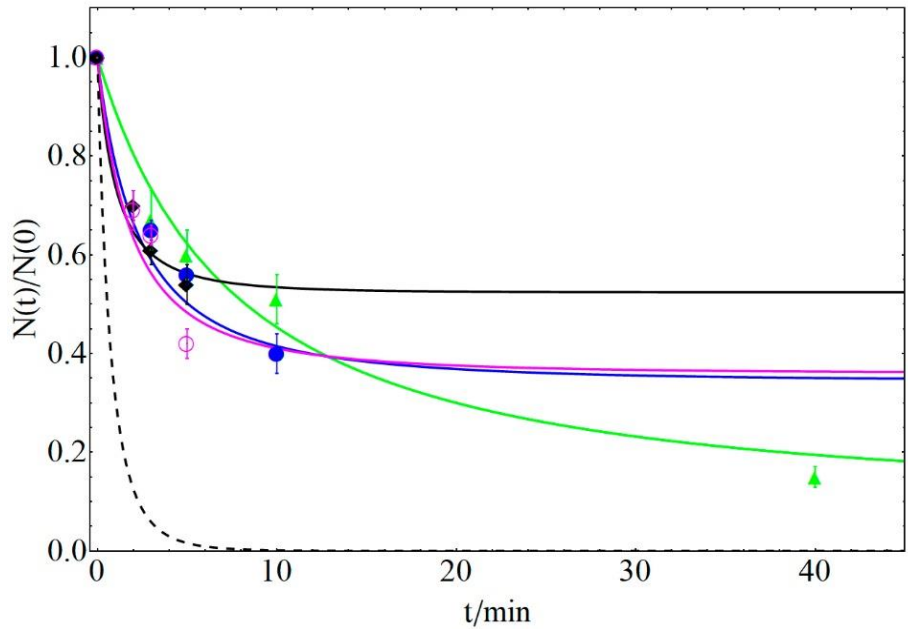

**Fig. 7** Comparison between the experimental $N(t)/N(0)$ data points (see Table 1) and the kinetic model. Parameter values employed in the numerical integration of eqn. 2 are $c = 3.06$ nm and $r^* = 3.58$ for 300 W and 0.5% $CH_4$ (green triangles), $r^* = 2.95$ for 600 W and 0.5% $CH_4$ (blue full circles), $r^* = 2.67$ for 800 W and 0.5% $CH_4$ (black full diamond), $r^* = 2.92$ nm for 600 W with 1% $CH_4$ (magenta open circles). The mean values of the growth rates are: 300 W, 0.5% $CH_4$, $k_g = 0.09$ nm min$^{-1}$; 600 W, 0.5% $CH_4$, $k_g = 0.4$ nm min$^{-1}$; 800 W, 0.5% $CH_4$, $k_g = 0.95$ nm min$^{-1}$; 600 W, 1% $CH_4$, $k_g = 0.5$ nm min$^{-1}$. Dashed line is the kinetics of etching in pure $H_2$ plasma estimated from Fig. 3 c (see text for details).

In the model a constant value of $c$, and therefore of $\sigma$, has been taken for all the curves, while $k_g$ was the only fitting parameter. Specifically, to estimate the $c$ parameter a mean value of diamond surface tension was employed, according to ref. [34][3]. The $k_g$ parameter is found to increase with the microwave plasma power, in agreement with the experimental evidence according to which the higher the power the larger the growth rate of the film. However, it is important to remember that $k_g$ is for the growth of the ND and can be different from that of a continuous thin film. In fact, the growth rate of thicker microcrystalline diamond (MCD) films was found to be greater than that of ultrathin NCD films (see Supplementary Material, Table S1 and Fig. S5).

In Fig. 8 we report the behavior of the density of seeds that survived the etching process as a function of the ratio $\frac{k_g}{K_e} > 0$. The computation is performed through the expression $\theta(k_g/K_e) = \int_{\frac{c}{\ln(K_g/K_e)}}^{\infty} f(x)dx$, using the lognormal distribution, $f(r)$, displayed in Fig. 2 and derived from PCS data.

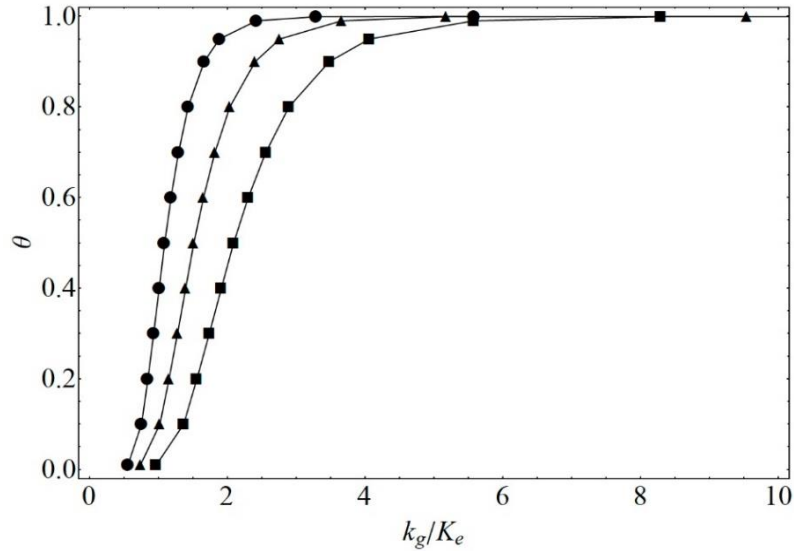

**Fig. 8** Fraction of the seed population that contribute to the film growth as a function of the $k_g/K_e$ ratio. Computations were performed using the $f(r)$ derived from PCS data (see Fig. 2) and for $c = 2$ nm (solid circle), $c = 2.5$ nm (solid triangle) and $c = 3.06$ nm (solid square).

From Fig. 8 ($c = 3.06$ nm curve) and the $k_g$ values obtained from the kinetic data (Fig. 7), we computed the $K_e$ and $K_g$ values reported in Fig. 9A. In fact, from the knowledge of the experimental density of seeds left on the surface ($\theta$) and the plot of Fig. 9A, it is possible to estimate the ratio of these rate coefficients. Besides, since $k_g = K_g - K_e$, the values of both these quantities can be finally obtained. In the figure, solid and open symbols refer to diamond deposition with 0.5% and 1% CH$_4$ in the feed gas, respectively. The net rates of growth of ND derived by the present model, $k_g$, are displayed in panel B of Fig. 9 (squares). In the same plot, the growth rates of the film at coalescence (Table 2) are also displayed as triangles; full symbols refer to 0.5% CH$_4$ and open symbols refer to 1% CH$_4$. The difference between the growth rates of seeds and NCD films at coalescence is attributable to the change of the crystalline habit of the seeds upon etching, where some crystal faces could be preferentially etched. Under this assumption, the crystal habits with higher $k_g$ are thought to be "selected" by the etching process; this implies an increase of the $k_g$ values as deposition proceeds. As far as the data at 600 W are concerned, the rate coefficients for etching, $K_e$, for 0.5% and 1% CH$_4$ are 0.27 and 0.33 nm min$^{-1}$ respectively. We point out that, differently from the $K_g$ quantities, these etching rate constants are very similar. Besides, these $K_e$ values are lower than that estimated for etching in pure hydrogen plasma (600 W) and equal to 0.4 nm/min. This rate has been computed through the present modeling at $K_g = 0$, by assuming that 99.9% of initial seeds were missing after 10 min H$_2$ plasma exposure (Fig. 3$c$). The corresponding etching kinetics is displayed in Fig. 8 as dashed line. We recall that the etching rate constant the model predicts in the absence of methane in the feed gas is a rough underestimate we tempted by using one experiment only (Fig. 3$c$). Anyway, its larger value compared to the $K_e$ constants at 600 W of Fig. 9A is reasonable, and attributable to a higher atomic hydrogen concentration in pure H$_2$ plasma discharge in that H atoms can be consumed in the reaction CH$_4$ + H → CH$_3$ + H$_2$ [31, 35].

---

[3] We have used $\sigma = 3.6$ J/m², hence $c = \frac{2\sigma V_m}{RT} = \frac{2 \times 3.6 \times 3.439 \cdot 10^{-6}}{8.31 \times 973} = 3.06 \cdot 10^{-9}$ m = 3.06 nm.

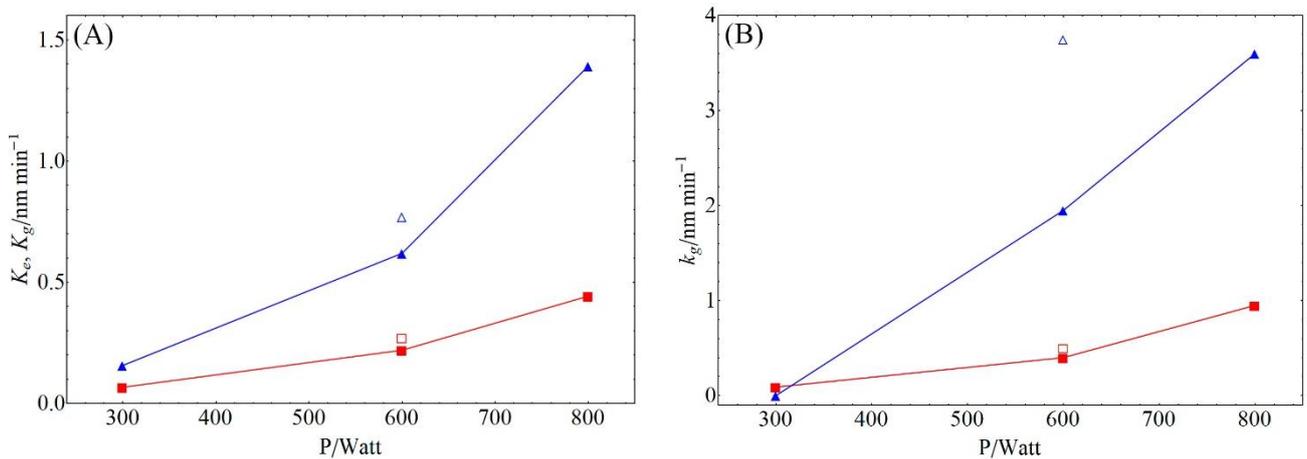

**Fig. 9** Panel (A): Behavior of the rate coefficients for growth ($K_g$, triangles) and etching ($K_e$, squares) as a function of microwave power, $P$, in the plasma. Solid symbols refer to deposition experiments carried out with 0.5% CH$_4$; open symbols refer to the deposition experiment with 1% CH$_4$ (600 W). Panel (B): Comparison between the net rate of growth ($k_g$) of a diamond film at coalescence (triangles) and the ND particles as derived by the kinetic model. Like in panel (A), solid and open symbols refer to depositions with 0.5% CH$_4$ and 1% CH$_4$, respectively.

Before concluding this section we stress that under similar experimental conditions (800 W, 0.5% $CH_4$, $p = 27$ Torr, $t = 120$ min) heterogeneous nucleation took place on the Si(100) substrate without any seeding pretreatment. As expected, we measured a nucleation density lower than $10^6$ cm$^{-2}$, i.e., more than five orders of magnitude less than seeding density (Fig. S6). The question therefore arises on how to reconcile the size dependent etching above discussed with the heterogeneous nucleation. In fact, heterogeneous nucleation implies the formation of critical clusters with radius surely lower than $r^*$ ($\cong 3$ nm, corresponding to $2 \times 10^4$ carbon atoms). This aspect can be tackled by considering, in the kinetic model, the effect of the interaction between the substrate and diamond. For this system, the chemical potential of C in ND is estimated on the basis of the Gibbs model of interface. The computation provides the expression (see Supplementary Material for details):

$$\mu_C = \mu_C(\infty) + \left(\frac{3}{2} - \frac{\beta}{2\sigma}\right)\frac{2\sigma v_m}{r}, \qquad (7)$$

where $\beta$ is the work of adhesion. An estimate of the radius dependent term in eqn. 7 can be obtained through the Born-Stern approach [36]. By setting $\beta \cong \varepsilon_{Si-C}$ and $2\sigma \cong \varepsilon_{C-C}$, $\varepsilon$ being the bonding energies between the stated couple of atoms in diamond (3.673 eV, ref. [37]) and SiC (4.6 eV, ref. [38]), we get $\left(\frac{3}{2} - \frac{\varepsilon_{Si-C}}{\varepsilon_{C-C}}\right) \cong 0.25$. Similar figure is attained by employing the work of adhesion, $\beta = 9.15$ J/m$^2$, recently calculated for Diamond (001)-Si(001) in ref. [39]. In this case we get $\left(\frac{3}{2} - \frac{\beta}{2\sigma_{cv}}\right) = 0.23$. The above computations imply a reduction of the $c$ term entering eqn. 2, with a parallel decrease of the $r^*$ value down to 0.69-0.75 nm [4] ($r^*$ was around 3 nm for seeds). It is worth mentioning that the size of the critical nucleus in diamond heterogeneous nucleation is still a debated topic. S.T. Lee and coworkers hypothesized that the critical size of the nucleus should be less than 2 nm ($r < 1$ nm), which was the size of the smallest diamond crystallite on silicon they could detect by high resolution transmission electron microscopy (HRTEM) [40]. Other Authors have estimated critical nuclei of few tens of atoms [41, 42]. The aim of this work is not to determine the critical size of the diamond nucleus, but rather to provide a rationale for what experimentalists have been observing since years. Equation 7 suggests that crystallites nucleated from the gas phase, and smaller than DND particles, may be stable toward etching under CVD conditions thanks to the stronger interaction with the substrate.

---

[4] A hemispherical diamond nucleus with radius 0.7 nm contains 120-130 carbon atoms.

## 4 Conclusions

For the first time a physical model to explain the DND seeds' gasification by monohydrogen under typical diamond CVD conditions has been proposed. This phenomenon has been depicted by introducing a size-dependent activation energy for the etching reaction. In the model, the stress due to particle curvature, according to the Young-Laplace equation, does increase the chemical potential of carbon atoms in DND, thus reducing the activation energy of the etching reaction. The approach predicts that seeds with a radius lower than a critical value, $r^*$, do not survive the CVD process. The $r^*$ quantity depends on *i)* the ratio of rate coefficients for growth ($K_g$) and etching ($K_e$), *ii)* surface tension of diamond and *iii)* deposition temperature. Under our experimental conditions $r^*$ varied between 2.7 and 3.6 nm and increased with microwave power when methane concentration and substrate temperature were kept constant. For a given size distribution function of seeds having minimum size smaller than $r^*$, the lower the $r^*$ value, the larger the fraction of seeds that will survive and grow.

**Appendix A. Supplementary data**

Supplementary material related to this article can be found online.

# Size dependent etching of nanodiamond seeds in the early stages of CVD diamond growth: Supplementary material

R. Salerno[1,2], B. Pede[1,2], M. Mastellone[2], V. Serpente[2], V. Valentini[2], A. Bellucci[2], D.M. Trucchi[2], F. Domenici[1], M. Tomellini[1*], R. Polini[1,2*]

[1] Dipartimento di Scienze e Tecnologie Chimiche, Università di Roma "Tor Vergata" and Consorzio INSTM RU "Roma Tor Vergata", Via della Ricerca Scientifica 1, Rome, 00133, Italy.

[2] Istituto di Struttura della Materia (ISM), Consiglio Nazionale delle Ricerche (CNR), Sez. Montelibretti, DiaTHEMA Lab, Via Salaria km 29.300, 00015 Monterotondo, Italy.

**I - Derivation of $f(r)$ from $f_v(r)$.**

The volume distribution gives the [*Volume of particles*] *vs* [*particle radius*]. We denote this distribution as $f_v(r)$, with $r$ being the equivalent diameter of particle. The volume occupied by particles with diameters in the range $(r, r + dr)$ reads

$$dV = f_v(r)dr. \tag{S1}$$

On the other hand, the volume $dV$ is given by $dV = \frac{4}{3}\pi r^3 dN(r)$ with $dN(r)$ number of particles with radius between $r, r + dr$. By defining the PDF for particle radius, $f(r)$, we get $dN(r) \propto f(r)dr$ that leads to the expression

$$f_v(r)dr \propto \frac{4}{3}\pi r^3 \, f(r)dr \, ,$$

from which

$$f(r) = C\left[f_v(r)/\left(\tfrac{4}{3}r^3\right)\right], \tag{S2}$$

with $C$ normalization constant.

**II – Additional SEM pictures and Raman spectra.**

Panels *(a)-(c)* of Fig. S1 present the plan-view of NCD films deposited using 600 W and 800 W microwave power (see Table 2 of the article). Fig. S1 *(d)* shows the cross section of the thin film obtained at 800 W, with average thickness of 52 nm.

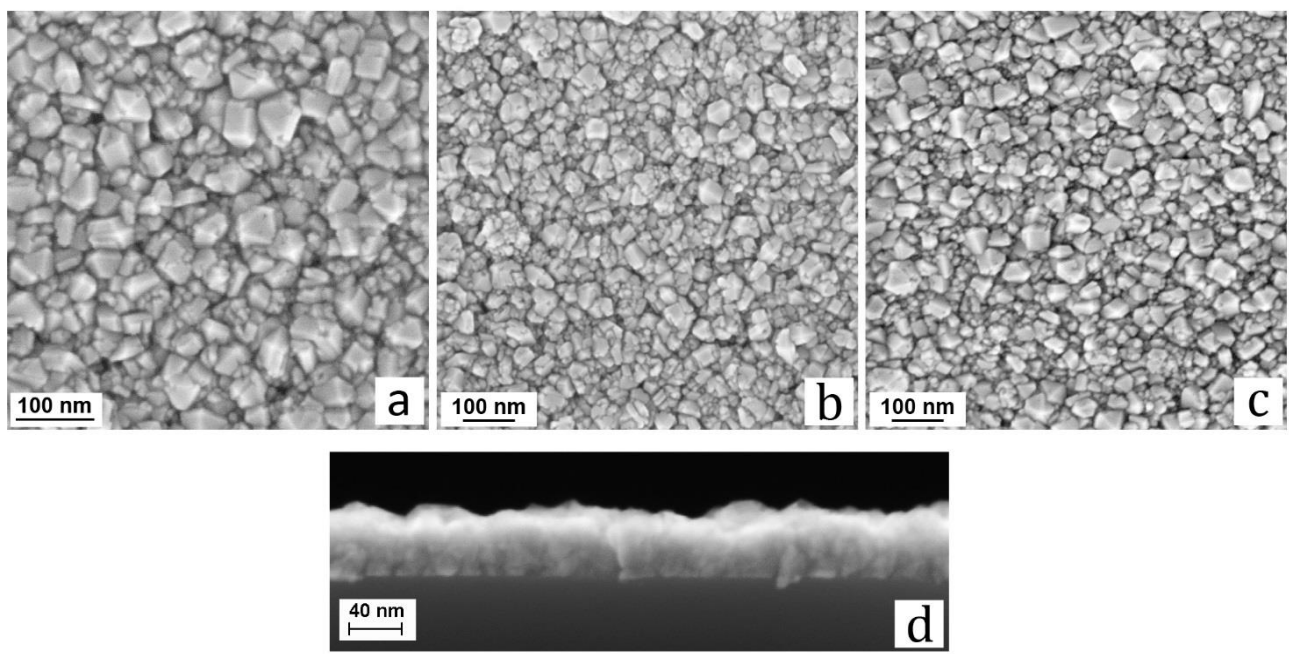

**Fig. S1** SEM micrographs of NCD films: *(a)* 600 W, 0.5% $CH_4$, 24.5 Torr; *(b)* 600 W, 1% $CH_4$, 24.5 Torr; *(c)* 800 W, 0.5% $CH_4$, 27 Torr; *(d)* cross section of the 52 nm thin film deposited at 800 W, 0.5% $CH_4$ (15 min).

Fig. S2 presents the Raman spectra of the NCD films of Fig. S1. The first peak at ~1125 cm$^{-1}$, labeled as $\nu_1$, can be assigned to *trans*-polyacetilene (sp² configuration) at the grain boundaries [1,2]. This attribution is confirmed by a second band at ~1450 cm$^{-1}$ ($\nu_3$). The diamond peak is located at 1333 cm$^{-1}$. Finally, the presence of graphitic carbon is witnessed by the G band at ~1550 cm$^{-1}$, and a barely visible D band at ~1350 cm$^{-1}$, attributable to disordered graphite.

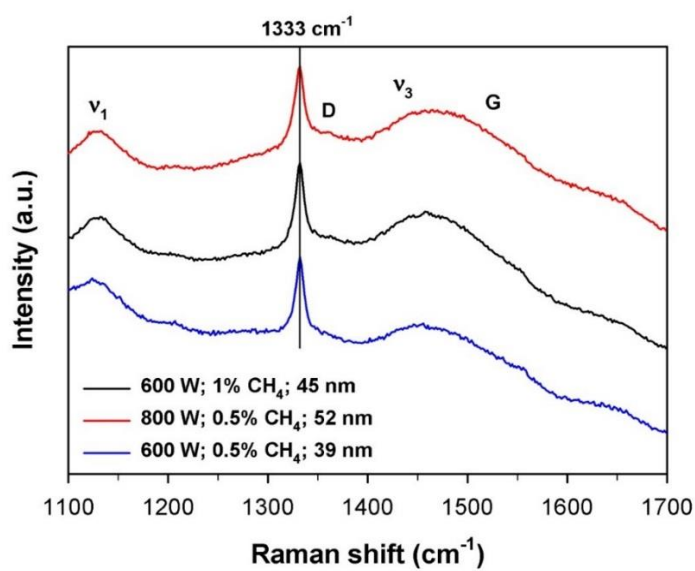

**Fig. S2** Raman spectra of NCD films.

Fig. S3 shows the ND particles remaining after 40 min MWCVD using 300 W microwave power and 0.5% methane in the feed gas (total $p$ = 8 Torr).

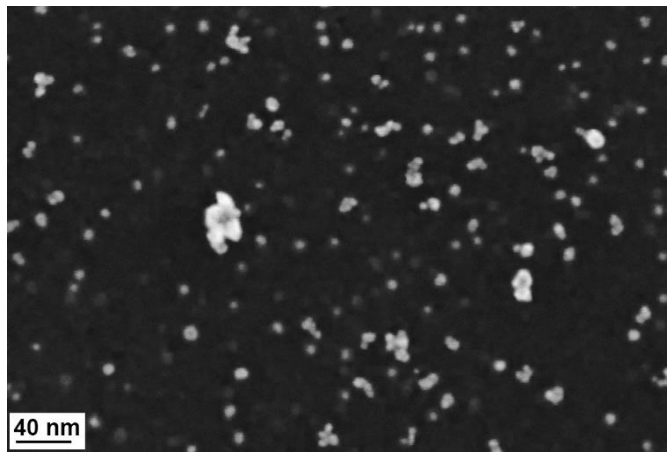

**Fig. S3** SEM micrograph of nanodiamond crystallites after 40 min CVD using 300 W, 0.5% $CH_4$ ($p$ = 8 Torr).

While the surface density of particles decreased (see Table 1 of the article), a very slow growth was observed after such a long CVD time, being several tens of nm the size of the largest crystallites. This is a consequence of both the small $K_g$ rate constant and rather similar $K_g$ and $K_e$ values at 300 W and 0.5% $CH_4$ (Fig. 9A). Under these CVD conditions, we got the smallest $K_g/K_e$ ratio at 300 W and 0.5% $CH_4$ and – consequently – the largest $r^*$ value through the relationship $r^* = \frac{c}{\ln(K_g/K_e)}$ (see manuscript). Nevertheless, diamond could grow more by using such a low plasma power and 8 Torr total pressure in the reactor chamber when we subjected the previously obtained 52 nm film shown in Fig. S2, panels *(c)* and *(d)* (15 min, 800 W, 0.5% $CH_4$), to a second CVD run for 300 min CVD, at 300 W, 0.5% $CH_4$; the film thickness did increase from 52 nm to 180 nm, with a rate of 0.43 nm/min (Fig. S4).

**III – Growth rates of NCD films at coalescence and of microcrystalline diamond (MCD) films.**

In Table 2 of the manuscript, we reported the deposition rates of ultrathin (~40-50 nm) NCD films prepared by employing 600 W and 800 W microwave powers. Such deposition rates were estimated by dividing the film thickness at coalescence by the CVD time. A very low growth rate was measured at 300 W, 0.5% $CH_4$, $p$ = 8 Torr: we could not observe the formation of a continuous ultrathin NCD film using these CVD parameters.

We performed longer deposition runs to grow thicker films.

As shown in Fig. S4, a thicker film could grow at 300 W and 0.5% $CH_4$ starting from a 52 nm film obtained under different CVD conditions. The growth rate of the "flat" film was 0.43 nm/min.

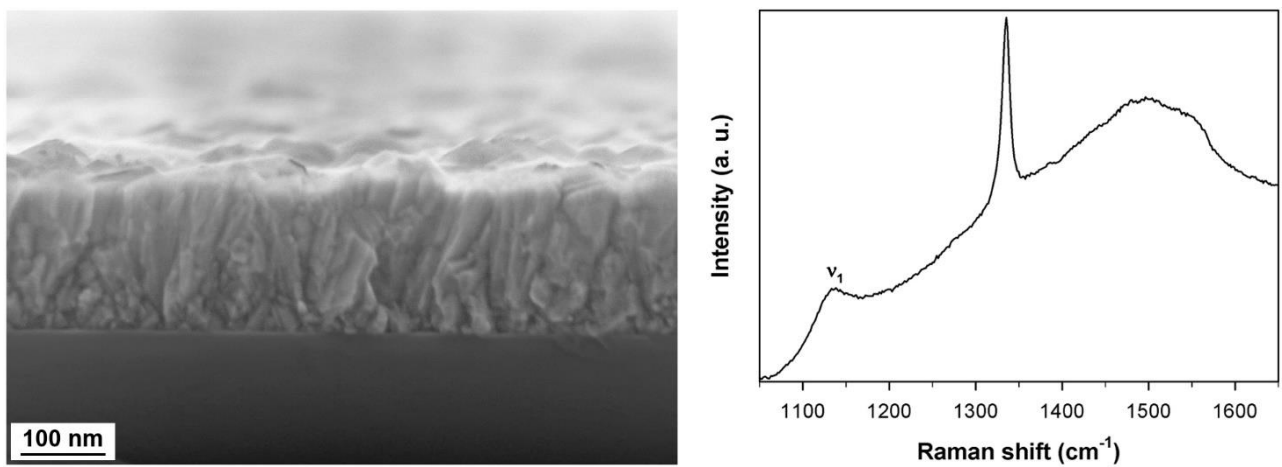

**Fig. S4** Cross section and Raman spectrum of a diamond film obtained by two-step MWCVD. The first deposition step was performed by exposing the as-seeded Si(100) substrate for 15 min to 0.5% $CH_4$, 800 W ($p$ = 27 Torr); the second step was run for 300 min at 300 W, 0.5% $CH_4$ ($p$ = 8 Torr). The overall thickness was 180 nm; therefore, the thickness increase, $\Delta z$, in the second step was $\Delta z$ = 180 – 52 = 128 nm, corresponding to a growth net rate of 0.43 nm/min.

These findings suggested us to determine the growth rates of thicker films deposited at 600 W and 800 W, and to compare said rates to the ones reported in Table 2 and referring to ultrathin NCD films growth.

| substrate | MW power (W) | $CH_4$ % | $p$ (Torr) | CVD time (min) | film thickness (nm) | deposition rate (nm/min) |
|---|---|---|---|---|---|---|
| seeded Si(100) | 300 | 0.5 | 8 | 40 | - | ~ 0 |
| 52 nm NCD film | 300 | 0.5 | 8 | 300 | 180 ± 10 | 0.43 ± 0.03 |
| seeded Si(100) | 600 | 0.5 | 24.5 | 20 | 39 ± 4 | 1.95 ± 0.2 |
| seeded Si(100) | 600 | 0.5 | 24.5 | 220 | 730 ± 30 | 3.3 ± 0.2 |
| seeded Si(100) | 600 | 1.0 | 24.5 | 12 | 45 ± 11 | 3.75 ± 0.9 |
| seeded Si(100) | 600 | 1.0 | 24.5 | 220 | 1020 ± 30 | 4.7 ± 0.2 |
| seeded Si(100) | 800 | 0.5 | 27 | 15 | 52 ± 10 | 3.5 ± 0.7 |
| seeded Si(100) | 800 | 0.5 | 27 | 220 | 1160 ± 60 | 5.3 ± 0.3 |
| seeded Si(100) | 800 | 1.0 | 27 | 10 | 52 ± 7 | 5.2 ± 0.7 |
| 52 nm NCD film | 800 | 1.0 | 27 | 210 | 1400 ± 50 | 6.5 ± 0.2 |

**Table S1** Thickness of the diamond films and corresponding deposition conditions and growth rates (substrate T = 700 °C).

The experimental details for ultrathin and thicker films deposition are summarized in Table S1. Substrate temperature was 700 °C for all the CVD runs. Film thickness values were measured by examining films' cross sections by SEM. The data of Table S1 are displayed in the plot of Fig. S5, and clearly demonstrate that deposition rates of thicker "flat" films are larger than those of ultrathin NCD films formed by impingement of growing seeds.

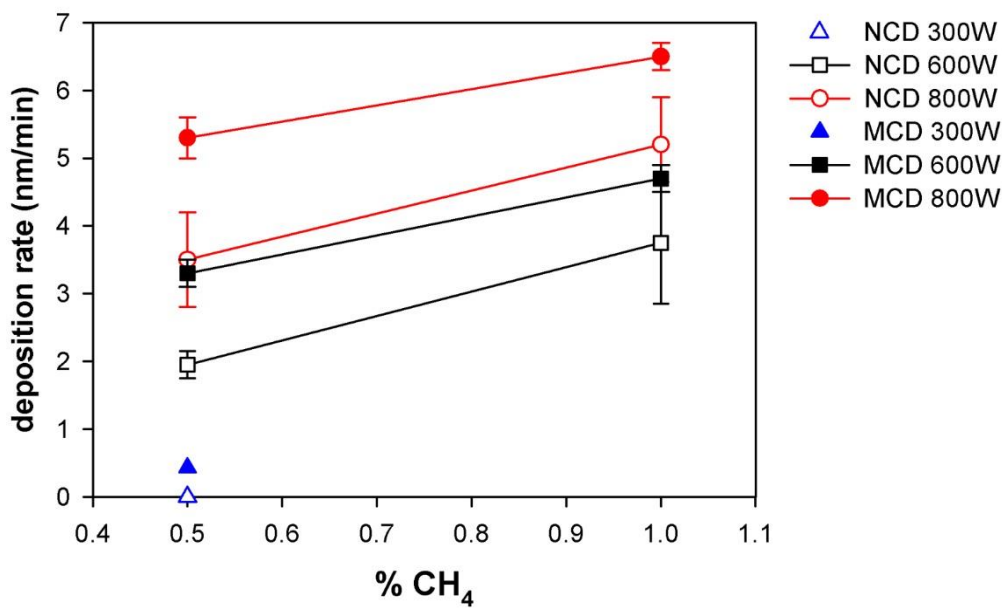

**Fig. S5** Different deposition rates of NCD (open symbols) and MCD (full symbols) films.

Fig. S6 shows the diamond crystals formed on a Si(100) substrate cleaned as detailed in the Experimental section (§ 3.1) but not seeded. CVD conditions were: 800 W microwave power, 0.5% $CH_4$ in the feed gas, 120 min deposition, substrate temperature T = 700 °C, total pressure $p$ = 27 Torr.

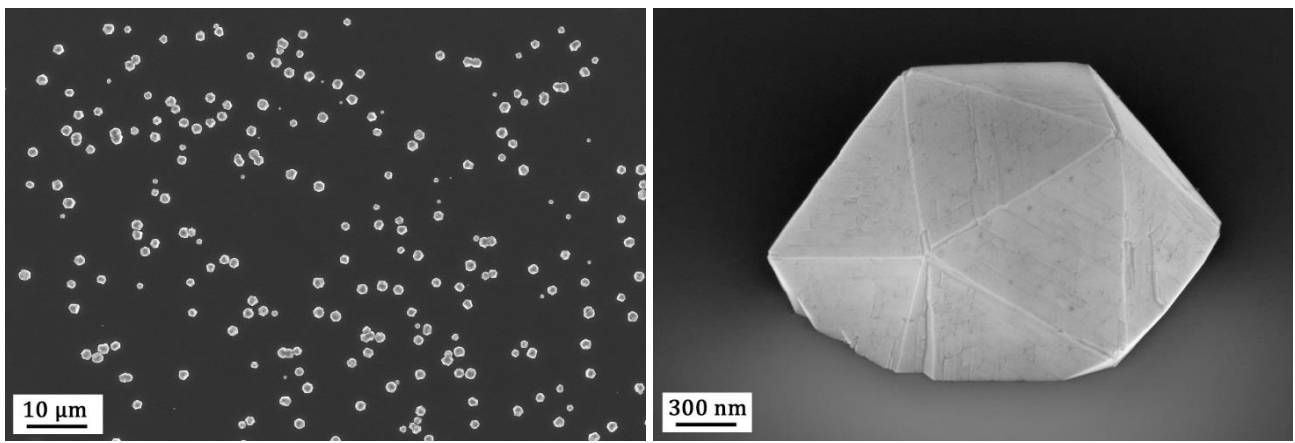

**Fig. S6** Diamond crystallites nucleated on virgin Si(100) substrate using the following CVD process conditions: 800 W microwave power, 0.5% $CH_4$ in the feed gas, $p$ = 27 Torr, $t$ = 120 min.

## IV – The case of heterogeneous nucleation.

The free energy of the system composed by a nucleus in contact with the substrate is given by

$$G = N_C \mu_C(\infty) + N_{S_i} \mu_{S_i}(\infty) + \sigma_{cs} A_{cs} + \sum_i \sigma_{cv,i} A_{cv,i} + \sigma_{sv} A_{sv}, \quad (S3)$$

where $\sigma_{kj}$ and $A_{kj}$ are, respectively, the excess free energy and area of the interface between $k$ and $j$ phases, where the subscripts $s$, $c$ and $v$ denote substrate, cluster (e.g., the sp³ carbon atoms cluster) and gas phase, respectively. In the equation, the sum is over the various faces of the nucleus. In the following, we make use of the average value of the excess free energy according to: $\sigma_{cv} = \frac{\sum_i \sigma_{cv,i} A_{cv,i}}{\sum_i A_{cv,i}} = \frac{\sum_i \sigma_{cv,i} A_{cv,i}}{A_{cv}}$. The $\sigma_{cv}$ term corresponds to the surface tension of diamond $\sigma$ (= 3.6 J/m²) we have used in the main text of the manuscript, including eqn. 7.

The chemical potential of C atoms reads

$$\mu_C = \frac{\partial G}{\partial N_C} = \mu_C(\infty) + (\sigma_{cs} - \sigma_{sv}) v_m \frac{\partial A_{cs}}{\partial V} + \sigma_{cv} v_m \frac{\partial A_{cv}}{\partial V}, \quad (S4)$$

with $V$ being the cluster volume and $v_m$ the molar volume of diamond. In eqn. S4 the equality $A_{cs} + A_{sv} = A_0$, $A_0$ being the total surface of the substrate, was also employed. Eqn. S4 is rewritten in terms of the work of adhesion $\beta = \sigma_{sv} + \sigma_{cv} - \sigma_{cs}$:

$$\mu_C = \mu_C(\infty) + (\sigma_{cv} - \beta) v_m \frac{\partial A_{cs}}{\partial V} + \sigma_{cv} v_m \frac{\partial A_{cv}}{\partial V}. \quad (S5)$$

To simplify the computation, we consider a hemispherical shape of the nucleus. In this case we obtain:

$$\mu_C = \mu_C(\infty) + \left(\frac{3}{2} - \frac{\beta}{2\sigma_{cv}}\right) \frac{2\sigma_{cv} v_m}{r}. \quad (S6)$$